\documentclass[3p,times,procedia]{elsarticle}
\usepackage{ecrc}
\usepackage{booktabs}
\usepackage{tikz}
\usepackage{subfig}
\usepackage{algorithm}
\usetikzlibrary{arrows,decorations.pathmorphing,backgrounds,positioning,fit,petri,shapes,automata}
\usepackage{amsmath}

\volume{00}

\firstpage{1}

\journalname{Procedia Computer Science}

\runauth{N. Tax et al.}

\jid{procs}

\usepackage{amssymb}

\usepackage{microtype}
\usepackage[figuresright]{rotating}
\usepackage{wrapfig}

\newcommand{\pre}[1]{\mbox{}^{\bullet}#1}

\newcommand{\fleche}{\longrightarrow}
\newcommand{\flsup}[1]{\stackrel{#1}{\fleche}}
\newcommand{\step}[1]{\flsup{#1}}           
\newcommand{\Lan}                 {\mathfrak{L}}
\newcommand{\N}                 {\mathbb{N}}
\newcommand{\E}                 {\mathcal{E}}

\newcommand{\indentiitem}{\setlength\itemindent{20pt}}
\newcommand{\indentiiitem}{\setlength\itemindent{40pt}}
\newcommand{\indentiiiitem}{\setlength\itemindent{60pt}}

\newcommand{\indentitem}{\setlength\itemindent{0pt}}
\begin{document}

\begin{frontmatter}



\dochead{20th International Conference on Knowledge Based and Intelligent Information and Engineering Systems}

\title{Log-based Evaluation of Label Splits for Process Models}


\author[a,b]{Niek Tax\corref{cor1}}
\author[a]{Natalia Sidorova}
\author[b]{Reinder Haakma}
\author[a]{Wil M. P. van der Aalst}

\address[a]{Eindhoven University of Technology, P.O. Box 513, Eindhoven, The Netherlands}
\address[b]{Philips Research, Prof. Holstlaan 4, 5665 AA Eindhoven, The Netherlands}

\begin{abstract}
Process mining techniques aim to extract insights in processes from event logs. One of the challenges in process mining is identifying interesting and meaningful event labels that contribute to a better understanding of the process. Our application area is mining data from smart homes for elderly, where the ultimate goal is to signal deviations from usual behavior and provide timely recommendations in order to extend the period of independent living. Extracting individual process models showing user behavior is an important instrument in achieving this goal. However, the interpretation of sensor data at an appropriate abstraction level is not straightforward. For example, a motion sensor in a bedroom can be triggered by tossing and turning in bed or by getting up. We try to derive the actual activity depending on the context (time, previous events, etc.). In this paper we introduce the notion of label refinements, which links more abstract event descriptions with their more refined counterparts. We present a statistical evaluation method to determine the usefulness of a label refinement for a given event log from a process perspective. Based on data from smart homes, we show how our statistical evaluation method for label refinements can be used in practice. Our method was able to select two label refinements out of a set of candidate label refinements that both had a positive effect on model precision.
\end{abstract}

\cortext[cor1]{Corresponding author. Tel.: +31-63-408-5760;}
\end{frontmatter}

\email{n.tax@tue.nl}


\vspace*{-5pt}


\section{Introduction}
Process mining is a fast growing discipline that brings together knowledge and techniques from computational intelligence, data mining, process modeling and process analysis~\cite{Aalst2011}. The process mining task is the automatic or semi-automatic analysis of events that are logged during process execution, where event records contain information on what was done, by whom, for whom, where, when, etc. Events are grouped into cases (process instances), e.g. per patient for a hospital log, or per insurance claim for an insurance company. An important task within process mining is \emph{process discovery}, which focuses on extracting interpretable models of processes from event logs. One of the attributes of the events is usually used as its label. These event labels are then used as transition/activity labels in the process models created by process discovery algorithms.

\begin{wraptable}{r}{8.3cm}
	\centering\scriptsize
	\caption{The corresponding smart home sensor event log with refined labels}
	\scalebox{0.75}{
		\hspace{-0.85cm}
		\begin{tabular}{ccccc|c}
			\toprule
			Id & Timestamp & Address & Sensor & Heart rate & Activity \\
			\midrule
			1 & \textcolor{gray}{03/11/2015} 02:45 & \textcolor{gray}{Mountain Rd. 7} & Bedroom motion & 74 & Tossing \& turning\\
			2 & \textcolor{gray}{03/11/2015} 03:23 & \textcolor{gray}{Mountain Rd. 7} & Bedroom motion & 72& Tossing \& turning\\
			3 & \textcolor{gray}{03/11/2015} 04:59 & \textcolor{gray}{Mountain Rd. 7} & Bedroom motion  & 71& Tossing \& turning\\
			4 & \textcolor{gray}{03/11/2015} 06:04 & \textcolor{gray}{Mountain Rd. 7} & Bedroom motion & 73 &Tossing \& turning\\
			5 & \textcolor{gray}{03/11/2015} 08:45 & \textcolor{gray}{Mountain Rd. 7} & Bedroom motion & 85 & Getting up\\
			6 & \textcolor{gray}{03/11/2015} 09:10 & \textcolor{gray}{Mountain Rd. 7} & Living room motion & 79 & Living room motion \\
			\dots & \textcolor{gray}{03/11/2015} \dots & \textcolor{gray}{Mountain Rd. 7} & \dots & \dots & \dots \\
			\midrule
			7 & \textcolor{gray}{03/12/2015} 01:01 & \textcolor{gray}{Mountain Rd. 7} & Bedroom motion & 73 & Tossing \& turning\\
			8 & \textcolor{gray}{03/12/2015} 03:13 & \textcolor{gray}{Mountain Rd. 7} & Bedroom motion & 75 & Tossing \& turning\\
			9 & \textcolor{gray}{03/12/2015} 07:24 & \textcolor{gray}{Mountain Rd. 7} & Bedroom motion & 74 &  Tossing \& turning\\
			10 & \textcolor{gray}{03/12/2015} 08:34 & \textcolor{gray}{Mountain Rd. 7} & Bedroom motion & 79 & Getting up\\
			11 & \textcolor{gray}{03/12/2015} 09:12 & \textcolor{gray}{Mountain Rd. 7} & Living room motion & 76 & Living room motion \\
			\dots & \textcolor{gray}{03/12/2015} \dots & \textcolor{gray}{Mountain Rd. 7} & \dots & \dots & \dots \\
			\midrule
			12 & \textcolor{gray}{03/13/2015} 00:45 &\textcolor{gray}{Mountain Rd. 7} & Bedroom motion & 75& Tossing \& turning\\
			13 & \textcolor{gray}{03/13/2015} 02:29 & \textcolor{gray}{Mountain Rd. 7} & Bedroom motion & 75 & Tossing \& turning\\
			14 & \textcolor{gray}{03/13/2015} 05:19 & \textcolor{gray}{Mountain Rd. 7} & Bedroom motion & 74 & Tossing \& turning\\
			15 & \textcolor{gray}{03/13/2015} 05:34 & \textcolor{gray}{Mountain Rd. 7} & Bedroom motion & 79& Tossing \& turning\\
			16 & \textcolor{gray}{03/13/2015} 05:39 & \textcolor{gray}{Mountain Rd. 7} & Bedroom motion & 77 & Tossing \& turning\\
			17 & \textcolor{gray}{03/13/2015} 08:37 & \textcolor{gray}{Mountain Rd. 7} & Bedroom motion & 79 & Getting up\\
			18 & \textcolor{gray}{03/13/2015} 08:52 & \textcolor{gray}{Mountain Rd. 7} & Living room motion & 78 &  Living room motion \\
			\dots & \textcolor{gray}{03/13/2015} \dots & \textcolor{gray}{Mountain Rd. 7} & \dots & \dots & \dots \\
			\midrule
			19 & \textcolor{gray}{03/14/2015} 03:41 & \textcolor{gray}{Mountain Rd. 7} & Bedroom motion & 75 & Tossing \& turning\\
			20 & \textcolor{gray}{03/14/2015} 05:00 & \textcolor{gray}{Mountain Rd. 7} & Bedroom motion & 74 & Tossing \& turning\\
			21 & \textcolor{gray}{03/14/2015} 08:52 & \textcolor{gray}{Mountain Rd. 7} & Bedroom motion & 75 & Getting up\\
			22 & \textcolor{gray}{03/14/2015} 09:30 & \textcolor{gray}{Mountain Rd. 7} & Living room motion & 74 &  Living room motion \\
			\dots & \textcolor{gray}{03/14/2015} \dots & \textcolor{gray}{Mountain Rd. 7} & \dots & \dots & \dots \\
			\midrule
			23 & \textcolor{gray}{03/15/2015} 02:11 & \textcolor{gray}{Mountain Rd. 7} & Bedroom motion & 77 & Tossing \& turning\\
			24 & \textcolor{gray}{03/15/2015} 02:34 & \textcolor{gray}{Mountain Rd. 7} & Bedroom motion & 76 & Tossing \& turning\\
			25 & \textcolor{gray}{03/15/2015} 08:35 & \textcolor{gray}{Mountain Rd. 7} & Bedroom motion & 79 & Getting up\\
			26 & \textcolor{gray}{03/15/2015} 08:57 & \textcolor{gray}{Mountain Rd. 7} & Living room motion & 77 & Living room motion \\
			\dots & \textcolor{gray}{03/15/2015} \dots & \textcolor{gray}{Mountain Rd. 7} & \dots & \dots & \dots \\
			\bottomrule
		\end{tabular}}
		\label{tab:evaluation_log}
		\vspace{-0.5cm}
	\end{wraptable}

Process mining takes its roots in the field of business process management, where the definition of labels for events is considered to be rather straightforward. In recent years, the application domain of process mining has broadened. A wide variety of event types can be used as input and analysis may be challenging. One of the most challenging application areas is \emph{LifeLogging}, which focuses on acquisition and analysis of personal daily life data. LifeLogs amongst others combine data collected through mobile phones, wearable devices, and/or smart home sensors. The emergence of LifeLogging tools and the resulting increase in availability of activity data enable a process-centric analysis of human behavior\cite{Sztyler2015}. The aim of process mining analysis on LifeLogging data is to find frequent activity patterns and represent them in a human interpretable process model. Such a process model could then also be used to detect deviations from one's regular behavior. Process mining in the human behavior application domain closely relates to the field of activity recognition, which aims to detect human activities from sensors and finding patterns between human activities \cite{Chen2012}. Process mining, however, aims to produce interpretable models that can provide insights by visually inspecting them. In contrast, most activity recognition techniques produce non-interpretable models.\looseness=-1

\begin{figure}
	\centering
	\scalebox{0.85}{
	\begin{tikzpicture}
	[node distance=1.11cm,
	on grid,>=stealth',
	bend angle=20,
	auto,
	every place/.style= {minimum size=4mm},
	every transition/.style = {minimum size = 8.5mm}
	]
	\node [place, tokens = 1] (p1) [label=below:$p_1$]{};
	\node [transition] (a) [label=center:{\tiny \parbox{20pt}{$Tossing$ $\&$ $turning$ $in$ $bed$}}, label=below:$t_1$, left = of p1] {}
	edge [pre, bend left] node[auto] {} (p1)
	edge [post, bend right] node[auto] {} (p1);
	\node [transition] (b) [label=center:{\tiny \parbox{20pt}{$Getting$ $out$ $of$ $bed$}}, label=below:$t_2$, right = of p1] {}
	edge [pre] node[auto] {} (p1);
	\node [place] (p2) [label=below:$p_2$, right = of b]{}
	edge [pre] node[auto] {} (b);
	\node [transition] (c) [label=center:{\tiny \parbox{20pt}{$Living$ $room$ $motion$}}, label=below:$t_3$, right = of p2] {}
	edge [pre] node[auto] {} (p2);
	\node [place] (p3) [label=below:$p_3$, right = of c]{}
	edge [pre] node[auto] {} (c);
	\node [transition] (d) [label=center:{\tiny \parbox{20pt}{$Kitchen$ $motion$}}, label=below:$t_4$, right = of p3] {}
	edge [pre] node[auto] {} (p3);
	\node [place] (p4) [label=above:$p_4$, below right = of d]{}
	edge [pre] node[auto] {} (d);
	\node [transition] (e) [label=center:{\tiny \parbox{20pt}{$Open$  $close$ $fridge$}}, label=above:$t_5$, right = of p4] {}
	edge [pre] node[auto] {} (p4);
	\node [place] (p6) [label=above:$p_6$, right = of e]{}
	edge [pre] node[auto] {} (e);
	\node [place] (p5) [label=above:$p_5$, above right = of d]{}
	edge [pre] node[auto] {} (d);
	\node [transition] (f) [label=center:{\tiny \parbox{20pt}{$Boil$ $water$}}, label=above:$t_6$, right = of p5] {}
	edge [pre] node[auto] {} (p5);
	\node [place] (p7) [label=above:$p_7$, right = of f]{}
	edge [pre] node[auto] {} (f);
	\node [transition] (g) [label=center:{\tiny \parbox{20pt}{$Living$ $room$ $motion$}}, label=below:$t_7$, below right = of p7] {}
	edge [pre] node[auto] {} (p7)
	edge [pre] node[auto] {} (p6);
	\node [place] (p8) [label=below:$p_8$, right = of g]{}
	edge [pre] node[auto] {} (g);
	
	\end{tikzpicture}}
	\caption{A Petri net derived from the event log in Table \ref{tab:evaluation_log}}
	\label{fig:example_petri_net}
	\vspace{-0.75cm}
\end{figure}
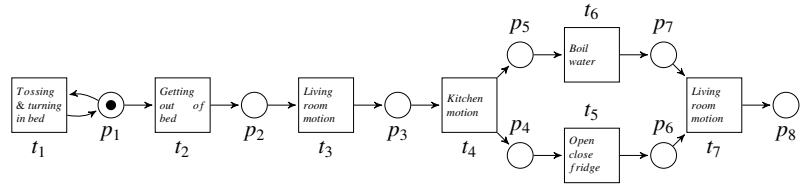

Imagine an elderly person of whom we want to discover a process model describing his/her daily behavior. Events are generated by sensors, either periodically (e.g. by a temperature sensor or heart rate monitor), or triggered by some activity (e.g. motion). Table~\ref{tab:evaluation_log} shows an example log obtained by fusing data from such sensors. The dots indicate that only a fraction of the logged events are shown. Assigning meaningful labels to these events is not straightforward. A \emph{Bedroom motion} event can be caused by different human activities, e.g. by \emph{Tossing \& turning} or by \emph{Getting up}. In some cases it is necessary to distinguish between \emph{Tossing \& turning} and \emph{Getting up}, for example when we aim to generate a timely reminder to take medication that needs to be taken before breakfast. Based on contextual information (e.g. a specific increase in heart rate, a time stamp, etc.), the distinction between the two types of activities might be identified, and each event with label \emph{Bedroom motion} can be refined into either \emph{Tossing \& turning} or \emph{Getting up}. The last column in Table~\ref{tab:evaluation_log} shows the desired event labels. Figure~\ref{fig:example_petri_net} shows a process model that can be deduced from such a log using existing process discovery techniques, like the ones from~\cite{Aalst2004,Weijters2011}.\looseness-1

Many relabelings of \emph{Bedroom motion} events are possible. Expert knowledge, data mining or machine learning techniques can be used to generate ideas for potential labeling functions. The goal of this labeling function is to give ``similar'' events the same label. However, similarity is a relative notion, \emph{so the initially chosen labeling function can be too abstract or too fine-grained to generate an informative process model}. Once a process discovery algorithm has been applied and a process model is obtained, one can assess whether the labeling function used on the original event log allowed the process discovery algorithm to discover an informative process model. However, it is computationally costly to apply process mining algorithms to multiple event logs generated from a single original event log using different event labeling functions with varying levels of abstraction. Therefore, we provide a statistical approach to evaluate label refinement usefulness in the context of process discovery that is based on significance testing of differences in event ordering relations.\looseness=-1

The Fodina \cite{Broucke2014} and the $\alpha^{*}$ \cite{Li2007} process discovery algorithms assume that there is one column in the event log that indicates the activity and refine this label based on a threshold of differentness on the event labels occurring directly before and after. In this paper we assume that the information what activity is performed is spread over multiple columns. We choose one column as primary activity column and refine the activity labels based on the other columns and temporal information. We validate whether a refinement makes sense from a process perspective by taking into account all temporal event information in the event log, using statistical testing and information gain. Evaluating splits based on information gain is a well-known approach in the area of decision tree learning \cite{Quinlan2014}, where ground truth labels are available in contrast to the label refinement setting. Label refinements draw similarities with automatic learning of ontologies \cite{Maedche2012} in the sense that both are concerned with inferring multiple levels of semantic interpretations from data. Ontology quality evaluation techniques \cite{Brank2005} can be used to evaluate (automatically inferred) ontologies, however these techniques are not process-centric, i.e., they do not take into account ordering relations between elements of the ontology in execution sequences.\looseness=-1

Section~\ref{sec:label_refinements} gives formal definitions of label refinements, process models, and related concepts. In Section \ref{sec:conceptual_notion_label_refinement}, we discuss when a label refinement is useful from a process mining perspective. A statistical method to evaluate the usefulness of a label refinement is described in Section~\ref{sec:calculate_label_split_quality}. In Section \ref{sec:evaluation_on_a_case_study} we discuss the results of the proposed method on a real life smart home data set. We draw conclusions in Section \ref{sec:conclusions}.\looseness-1

\section{Label Refinements \& Process Models}
\label{sec:label_refinements}
In this section we introduce the notions related to event logs and relabeling functions for traces and then define the notions of refinements and abstractions. We also introduce the Petri net process model notation.

We use the usual sequence definition, and denote a sequence by listing its elements, e.g. we write $\langle a_1,a_2,\dots,a_{n} \rangle$ for a (finite) sequence $s:\{1,\dots,n\}\to A$ of elements from some alphabet $A$, where $s(i)=a_i$ for any $i \in \{1,\dots,n\}$. The length of a sequence $s:\{1,\dots,n\}\to A$ is $|s|=n$; $s_1 s_2$ denotes the concatenation of sequences $s_1$ and $s_2$. A \emph{language} $\Lan$ over an alphabet $A$ is a set of sequences over $A$. $\Lan^p$ is the prefix closure of a language $\Lan$ (with $\Lan\subseteq\Lan^p$).

An event is the most elementary element of an event log. Let $\mathcal{I}$ be a set of event identifiers, $\mathcal{T}$ be a set of timestamps, and $\mathcal{A}_1 \times \dots \times \mathcal{A}_{n}$ be an attribute domain consisting of $n$ attributes (e.g. resource, activity name, cost, etc.), each of a certain type. An event is a tuple $e=(i,t,a_1,\dots,a_{n})$, with $i\in\mathcal{I}$, $t\in \mathcal{T}$, and $(a_1, \dots, a_{n})\in \mathcal{A}_1 \times \dots \times \mathcal{A}_{n}$. The \emph{event label} of an event is the attribute set $(a_1\dots,a_n)$; $e_i$, $e_t$ and $e_a$ respectively denote the identifier, the timestamp and label of event $e$.  $\mathcal{E}=\mathcal{I}\times\mathcal{T}\times \mathcal{A}_1 \times \dots \times \mathcal{A}_{n}$ is a universe of events over $\mathcal{A}_1, \dots, \mathcal{A}_{n}$.
The lines of Table \ref{tab:evaluation_log}, where we do not consider the \emph{activity} column for now, are events from an event universe over the event attributes \emph{sensor}, \emph{address}, and \emph{heart rate}.\looseness=-1

Events are often considered in the context of other events. We call $E\subseteq\mathcal{E}$ an \emph{event set}, if $E$ does not contain any events with the same event identifier. The events in Table \ref{tab:evaluation_log} together form an event set. A \emph{trace} $\sigma$ is a finite sequence formed by the events from an event set $E\subseteq{\mathcal{E}}$ that respects the time ordering of events, i.e. for all $k,m\in\N$, $1\leq k < m \leq |E|$, we have: $\sigma(k)_t\leq \sigma(m)_t$. We define the \emph{universe of traces} over event universe $\mathcal{E}$, denoted $\Sigma(\mathcal{E})$, as the set of all possible traces over $\mathcal{E}$. We omit $\mathcal{E}$ in $\Sigma(\mathcal{E})$ and use the shorter notation $\Sigma$ when the event universe is clear from the context.\looseness=-1 

Often it is useful to partition an event set into smaller sets in which events belong together according to some criterion. We might for example be interested in discovering the typical behavior of households over the course of a day. In order to do so, we can e.g.~group together events with the same \emph{address} and the same day-part of the \emph{timestamp}, as indicated by the horizontal lines in Table \ref{tab:evaluation_log}. For each of these event sets, we can construct a trace; time stamps define the ordering of events within the trace. For events of a trace having the same time stamps, an arbitrary ordering can be chosen within a trace.

An \emph{event partitioning function} is a function $ep: \mathcal{E} \to T_{id}$ that defines the partitioning of an arbitrary set of events $E\subseteq\mathcal{E}$ from a given event universe $\mathcal{E}$ into event sets $E_1,\ldots,E_j,\ldots$ where each $E_j$ is the maximal subset of $E$ such that for any $e_1, e_2\in E_j$, $ep(e_1)= ep(e_2)$; the value of $ep$ shared by all the elements of $E_j$ defines the value of the \emph{trace attribute} $T_{id}$. Note that complex, multidimensional trace attributes are also possible, i.e. a combination of the name of the person performing the event activity and the date of the event, so that every trace contains activities of one person during one day. The event sets obtained by applying an event partitioning can be transformed into traces (respecting the time ordering of events).


An event log $L$ is a finite set of traces $L \subseteq \Sigma(\mathcal{E})$. $A_L\subseteq\mathcal{A}_1 \times \dots \times \mathcal{A}_{n}$ denotes the \emph{alphabet of event labels} that occur in log $L$. The traces of a log are often transformed before doing further analysis: very detailed but not necessarily informative event descriptions are transformed into some \emph{informative} and \emph{repeatable} labels. For the labels of the log in Table~\ref{tab:evaluation_log}, the heart rate values can be abstracted to \textit{low, normal}, and \textit{high} or the label can be redefined to a subset of the event attributes. Next to that, if the event partitioning function maps each event from Table~\ref{tab:evaluation_log} to its address and the day-part of the timestamp, these attributes (indicated in gray) become the trace attribute and can safely be removed from individual events. The new label is then defined as a combination of the sensor and abstracted heart rate values.

After this relabeling step, some traces of the log can become identically labeled (the event id's would still be different). The information about the number of occurrences of a sequence of labels in an event log is highly relevant for process mining, since it allows differentiating between the main stream behavior of a process (frequently occurring behavioral patterns) and exceptional behavior.

Let $\Sigma(\mathcal{E})$ and $\Sigma'(\mathcal{E'})$ be two universes of traces defined over event universes $\mathcal{E},\mathcal{E'}$. A function $l: \Sigma \to \Sigma'$ is a \emph{trace relabeling function} if for all traces $\sigma,\gamma \in \Sigma$ such that if $\sigma$ is a prefix of $\gamma$, $l(\sigma)$ is a prefix of or equal to $l(\gamma)$. We lift $l$ to event logs: for $L \subseteq \Sigma$, the relabeling $l(L)$ is defined as $\{l(\sigma)|\sigma\in L\}$.

Often, relabeling functions are defined using a more narrow approach: first defining an event relabeling function and then lifting that function to traces. In the context of business processes, event relabeling functions are mostly mere projections of events on the values of a single attribute, such as \textit{activity name}. We consider a more general definition to allow for history-dependent interpretation of events, which is necessary in the context of LifeLogging. Prefix preservation requirement is necessary to allow for logging, compliance checking and other forms of analysis performed at run time.\looseness-1

Let $\Sigma$, $\Sigma_1$, and $\Sigma_2$ be trace universes over $\E,\E_1,\E_2$ respectively with $\E,\E_1,\E_2$ being pairwise different. Let $l_1: \Sigma \to \Sigma_1$ and $l_2: \Sigma \to \Sigma_2$ be trace relabeling functions. Relabeling function $l_1$ is a \emph{refinement} of relabeling function $l_2$, denoted by $l_1\preceq l_2$, iff $\forall_{\sigma_1,\sigma_2\in \Sigma}:l_1(\sigma_1)=l_1(\sigma_2)\implies l_2(\sigma_1)=l_2(\sigma_2)$; $l_2$ is then called an \emph{abstraction} of $l_1$. We call a refinement $l_1$ of $l_2$ a \emph{strict} refinement, denoted by $l_1\prec l_2$, when $\exists_{\sigma_1,\sigma_2\in \Sigma}: l_1(\sigma_1)\neq l_1(\sigma_2)\wedge l_2(\sigma_1)=l_2(\sigma_2)$. We call refinement $l_1$ of $l_2$ an \emph{equal length} refinement, denoted by $l_1\preceq^= l_2$,when $\forall{\sigma\in \Sigma}:|l_1(\sigma)|=|l_2(\sigma)|$.

Let $\Sigma,\Sigma_1$ be trace universes over $\E,\E_1$ respectively,  $l: \Sigma \to \Sigma_1$ a trace relabeling function, and $\Lan_1$ be a language $\Lan_1 \subseteq \Sigma_1$ over $\E_1$. \emph{Trace concretization} $l^{-1}:\Sigma_1\to2^{\Sigma}$ is a function defined as $l^{-1}(\sigma_1)=\{\sigma\in\Sigma|l(\sigma)=\sigma_1\}$, for each $\sigma_1\in\Sigma_1$. \emph{Language concretization} of $\Lan_1$ is language $l^{-1}(\Lan_1)=\cup_{\sigma_1\in \Lan_1}l^{-1}(\sigma')$.

The goal of process discovery is to discover a process model that represents the behavior seen in an event log. A frequently used process modeling notation in the process mining field is the Petri net~\cite{Reisig1998}. Petri nets are directed bipartite graphs consisting of transitions and places, connected by arcs. Transitions represent activities, while places represent the enabling conditions of transitions. Labels are assigned to transitions to indicate the type of activity that they model. A special label $\tau$ is used to represent invisible transitions, which are only used for routing purposes and not recorded in the execution log.

A \emph{labeled Petri net} $N=\langle P,T,F,A_M,\ell\rangle$ is a tuple where $P$ is a finite set of places, $T$ is a finite set of transitions such that $P \cap T = \emptyset$,  $F \subseteq (P \times T) \cup (T \times P)$ is a set of directed arcs, called the flow relation, $A_M$ is an alphabet of labels representing activities, with $\tau \notin A_M$ being a label representing  invisible events, and $\ell:T\rightarrow A_M\cup \{\tau\}$ is a labeling function that assigns a label to each transition. For a node $n \in (P \cup T)$ we use $\bullet n$ and $n \bullet$ to denote the set of input and output nodes of $n$, defined as $\bullet n =\{n'|(n',n)\in F\}$ and $n \bullet =\{n|(n,n')\in F\}$. An example of a Petri net can be seen in Figure \ref{fig:example_petri_net}, where circles represent places and squares represent transitions. 

A state of a Petri net is defined by its \emph{marking} $M \in \mathbb{N}^{P}$ being a multiset of places. A marking is graphically denoted by putting $M(p)$ tokens on each place $p\in P$. A pair $(N,M)$ is called a marked Petri net. State changes occur through transition firings. A transition $t$ is enabled (can fire) in a given marking $M$ if each input place $p\in \pre{t}$ contains at least one token. Once a transition fires, one token is removed from each input place of $t$  and one token is added to each output place of $t$, leading to a new marking $M'$ defined as $M'=M-\bullet t+t\bullet$.
A firing of a transition $t$ leading from marking $M$ to marking $M'$ is denoted as $M \step{\ell(t)} M'$. $M_1 \step{\ell(\sigma)} M_2$ indicates that $M_2$ can be reached from $M_1$ through a firing sequence $\sigma'\in {A_M}^*$. Many process modeling notations have formal executional semantics and define a \emph{language of accepting traces} $\Lan$. For Petri net $N_2$ in Figure \ref{fig:precision}, $\Lan(N_2)=\{\langle\textit{Bedroom motion},\textit{Livingroom Motion}\rangle,\langle\textit{Bedroom motion},\textit{Bedroom motion},\textit{Livingroom Motion}\rangle,\langle\textit{Bedroom motion},\\\dots,\textit{Bedroom motion},\textit{Livingroom Motion}\rangle\}$.

\section{On the Quality of Label Refinements for Process Mining}
\label{sec:conceptual_notion_label_refinement}

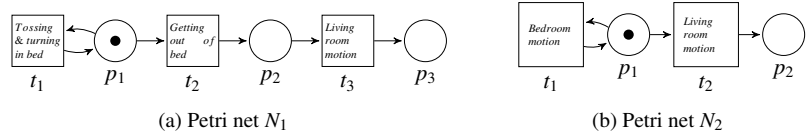
\begin{figure}[t]
	\centering
	\subfloat[Petri net $N_1$]{
		\centering
		\scalebox{0.85}{
		\begin{tikzpicture}
		[node distance=1.2cm,
		on grid,>=stealth',
		bend angle=20,
		auto,
		every place/.style= {minimum size=6.5mm},
		every transition/.style = {minimum size = 8mm}
		]
		\node [place, tokens = 1] (p1) [label=below:$p_1$]{};
		\node [transition] (a) [label=center:{\tiny \parbox{20pt}{$Tossing$ $\&$ $turning$ $in$ $bed$}}, label=below:$t_1$, left = of p1] {}
		edge [pre, bend left] node[auto] {} (p1)
		edge [post, bend right] node[auto] {} (p1);
		\node [transition] (b) [label=center:{\tiny \parbox{20pt}{$Getting$ $out$ $of$ $bed$}}, label=below:$t_2$, right = of p1] {}
		edge [pre] node[auto] {} (p1);
		\node [place] (p2) [label=below:$p_2$, right = of b]{}
		edge [pre] node[auto] {} (b);
		\node [transition] (c) [label=center:{\tiny \parbox{20pt}{$Living$ $room$ $motion$}}, label=below:$t_3$, right = of p2] {}
		edge [pre] node[auto] {} (p2);
		\node [place] (p3) [label=below:$p_3$, right = of c]{}
		edge [pre] node[auto] {} (c);
		\end{tikzpicture}
	}}
	\qquad
	\subfloat[Petri net $N_2$]{
		\centering
		\scalebox{0.85}{
		\begin{tikzpicture}
		[node distance=1.2cm,
		on grid,>=stealth',
		bend angle=20,
		auto,
		every place/.style= {minimum size=6.5mm},
		every transition/.style = {minimum size = 10mm}
		]
		\node [place, tokens = 1] (p1) [label=below:$p_1$]{};
		\node [transition] (a) [label=center:{\tiny \parbox{20pt}{$Bedroom$ $motion$}}, label=below:$t_1$, left = of p1] {}
		edge [pre, bend left] node[auto] {} (p1)
		edge [post, bend right] node[auto] {} (p1);
		\node [transition] (b) [label=center:{\tiny \parbox{20pt}{$Living$ $room$ $motion$}}, label=below:$t_2$, right = of p1] {}
		edge [pre] node[auto] {} (p1);
		\node [place] (p2) [label=below:$p_2$, right = of b]{}
		edge [pre] node[auto] {} (b);
		\end{tikzpicture}
	}}
	\caption{Petri nets discovered from two event logs obtained from the same event set with different relabeling functions.}
	\vspace{-0.5cm}
	\label{fig:precision}
\end{figure}


\begin{wrapfigure}{R}{6.3cm}
	\vspace{-0.65cm}
	\centering
	\scalebox{0.71}{
		\hspace{-1.2cm}
		\begin{tikzpicture}
		[node distance=1cm,
		on grid,>=stealth',
		bend angle=10,
		auto,
		every transition/.style = {align=center, minimum size = 5mm}
		]
		\node [transition] (a) [] {\small{Event Set $E$}};
		\node [transition] (b) [below = of a] {\small{Event Log $L$}}
		edge [pre] node[auto] {\scriptsize{Trace attribute}} (a);
		\node [transition] (c) [below left = 1cm and 3cm of b]{\small{Event Log $L_1$=$l_1(L)$}}
		edge [pre] node[anchor=east] {\scriptsize{Relabeling function $l_1$}} (b);
		\node [transition] (d) [below right = 1cm and 3cm of b]{\small{Event Log $L_2$=$l_2(L)$}}
		edge [pre] node[anchor=west] {\scriptsize{Relabeling function $l_2$}} (b)
		edge [dashed] node[auto] {\scriptsize{$l_1\preceq l_2$}} (c);
		\node [transition] (e) [below = of c]{\small{Process Model $N_1$}}
		edge [pre] node[auto] {\scriptsize{Process Discovery}} (c);
		\node [transition] (f) [below = of d]{\small{Process Model $N_2$}}
		edge [pre] node[anchor=west] {\scriptsize{Process Discovery}} (d);
		\node [transition] (g) [below = of e]{\small{Language $\Lan_1$}}
		edge [pre] node[anchor=east] {\scriptsize{Generates}} (e);
		\node [transition] (h) [below = of f]{\small{Language $\Lan_2$}}
		edge [pre] node[anchor=west] {\scriptsize{Generates}} (f);
		\node [transition] (i) [below = 1.5cm of g]{\small{Language $l_1^{-1}(\Lan_1)$}}
		edge [pre] node[auto] {\scriptsize{$\begin{array}{r}
				\text{Language}\\
				\text{Concretization}\\
				\end{array}$}} (g)
		edge [dashed, bend right] node[anchor=west] {\scriptsize{Fitness}} (b);
		\node [transition] (j) [below = 1.5cm of h]{\small{Language $l_2^{-1}(\Lan_2)$}}
		edge [pre] node[anchor=west] {\scriptsize{$\begin{array}{l}
				\text{Language}\\
				\text{Concretization}\\
				\end{array}$}} (h)
		edge [dashed] node {$\supseteq$}(i)
		edge [dashed, bend left] node[anchor=east] {\scriptsize{Fitness}} (b);
		\end{tikzpicture}}
	\caption{Comparing two event relabeling functions}
	\label{fig:refinement}
\end{wrapfigure}
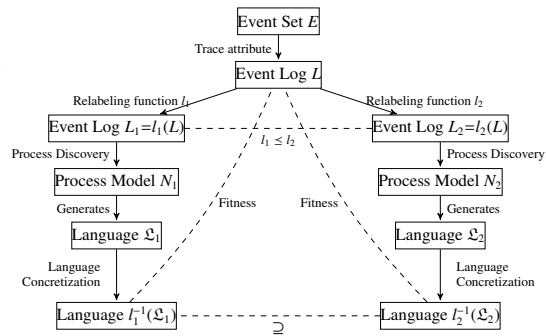

\emph{Process discovery algorithms} discover a process model based on an event log, where event labels are obtained by applying an event relabeling function to an original log. The main quality metrics discovered process models are fitness, precision, generalization and simplicity~\cite{Aalst2011}. \emph{Fitness} represents the share of the behavior seen in the log that is allowed by the process model. \emph{Precision} aims at narrowing the set of traces that belong to the language of the discovered process model, but was not observed in the event log. 
\emph{Generalization} aims at preventing overfitting, and \emph{simplicity} measures the ``understandability'' and ``well-structuredness'' of models.\looseness=-1

Intuitively, an event relabeling function is better than another one if it improves the quality of the discovered model along these quality dimensions. However, the quality metrics are currently defined in such a way that only results of discovery algorithms applied to the very same log can be compared, while two different relabeling functions produce logs with different event labels. The Petri net $N_1$ in Figure~\ref{fig:precision} has perfect precision and fitness for the event log with labels as shown in the refined label column of Table~\ref{tab:evaluation_log}. At the same time, Petri net $N_2$ has perfect fitness and precision for the event log with labels as in the sensor column of Table~\ref{tab:evaluation_log}. However, Petri net $N_1$ is useful for the purpose of sending a reminder message to take medicines after getting up, while Petri net $N_2$ is not. This suggests that Petri net $N_1$ is more precise than $N_2$, but only with respect to the original log.
Thus we have to make the comparison in the context of the original log. Suppose we have a set of events $E$, which is part of some universe of events $\E$. We choose a case identifier and build an event log $L$ from $E$. Then we choose relabeling functions $l_1$ and $l_2$ with $l_1 \prec l_2$ and obtain $L_1=l_1(L)$ and $L_2=l_2(L)$ (see Figure~\ref{fig:refinement}).
Applying process discovery to $L_1$ and $L_2$ results in two process models, which respectively accept languages $\Lan_1$ and $\Lan_2$. These languages cannot be compared directly, since they contain traces consisting of different event labels. Precision metrics look at ``redundant'' traces in the mined models with respect to the log used as input for the discovery algorithm (see e.g.~\cite{Munoz2010,Rozinat2008}).
Using the inverse functions $l^{-1}_1$, $l^{-1}_2$, every trace of $\Lan_1$ and $\Lan_2$ can be mapped to a set of traces built from the events from $\E$. Taking the union of the sets obtained with $l^{-1}_1$, $l^{-1}_2$ over the traces of the languages, we obtain comparable languages and can conclude whether the relabeling function results in a model that is more precise with respect to the original log.

Fitness and simplicity of the models depend mostly on the performance of the process discovery algorithm, and not on the choice of the relabeling function. Precision defined in terms of events of the original universe $\E$ of events is however highly dependent on the appropriateness of the relabeling function: choosing a more refined relabeling function can increase the precision by eliminating the behavior that would be allowed in the model discovered with a more abstract relabeling function. Generalization can potentially suffer as the result of a higher precision.
\vspace{-0.4cm}
\subsection{Label Refinement Quality}
\label{ssec:split_quality_conceptual}
The comparison of the languages generated by models is not feasible due to its complexity; for many classes of process models, including Petri nets, the problem of language inclusion is just not decidable. Therefore, we need a different, practical approach to deciding on the usefulness of a relabeling function refinement. We start with discussing the usefulness by comparing the discovered models.

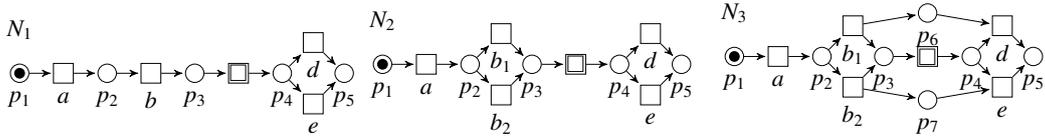
\begin{figure}[t]
	\vspace{-0.2cm}
	\centering
	\scalebox{0.9}{
	\subfloat{
		\hspace{-0.3cm}
		\begin{tikzpicture}
		[node distance=0.64cm,
		on grid,>=stealth',
		bend angle=17.5,
		auto,
		every place/.style= {minimum size=2.7mm},
		every transition/.style = {minimum size = 2.9mm}
		]
		
		\node [place, tokens = 1] (p1) [label = below:$p_1$]{};
		
		\node[] (N1) [above = of p1]{$N_1$};	
		
		\node [transition] (a) [label = below:$a$, right = of p1]{}
		edge [pre] node[auto] {} (p1);
		
		\node [place] (p2) [label = below:$p_2$, right = of a]{}
		edge [pre] node[auto] {} (a);
		
		\node [transition] (b) [label = below:$b$, right = of p2]{}
		edge [pre] node[auto] {} (p2);
		
		\node [place] (p3) [label = below:$p_3$, right = of b]{}
		edge [pre] node[auto] {} (b);
		
		\node [transition] (c) [label = center:$\square$, right = of p3]{}
		edge [pre] node[auto] {} (p3);
		
		\node [place] (p4) [label = below:$p_4$, right = of c]{}
		edge [pre] node[auto] {} (c);
		
		\node [transition] (d) [label = below:$d$, above right = of p4]{}
		edge [pre] node[auto] {} (p4);
		
		\node [transition] (e) [label = below:$e$, below right = of p4]{}
		edge [pre] node[auto] {} (p4);
		
		\node [place] (p5) [label = below:$p_5$, below right = of d]{}
		edge [pre] node[auto] {} (d)
		edge [pre] node[auto] {} (e);
		
		\end{tikzpicture}
	}
	\subfloat{
		\hspace{-0.3cm}
		\begin{tikzpicture}
		[node distance=0.64cm,
		on grid,>=stealth',
		bend angle=17.5,
		auto,
		every place/.style= {minimum size=2.7mm},
		every transition/.style = {minimum size = 2.9mm}
		]
		\node [place, tokens = 1] (p1l) [label = below:$p_1$]{};
		\node[] (N2) [above = of p1l]{$N_2$};	
		\node [transition] (al) [label = below:$a$, right = of p1l]{}
		edge [pre] node[auto] {} (p1l);
		
		\node [place] (p2l) [label = below:$p_2$, right = of al]{}
		edge [pre] node[auto] {} (al);
		
		\node [transition] (b1l) [label = below:$b_1$, above right = of p2l]{}
		edge [pre] node[auto] {} (p2l);
		
		\node [transition] (b2l) [label = below:$b_2$, below right = of p2l]{}
		edge [pre] node[auto] {} (p2l);
		
		\node [place] (p3l) [label = below:$p_3$, below right = of b1l]{}
		edge [pre] node[auto] {} (b1l)
		edge [pre] node[auto] {} (b2l);
		
		\node [transition] (cl) [label = center:$\square$, right = of p3l]{}
		edge [pre] node[auto] {} (p3l);
		
		\node [place] (p4l) [label = below:$p_4$, right = of cl]{}
		edge [pre] node[auto] {} (cl);
		
		\node [transition] (dl) [label = below:$d$, above right = of p4l]{}
		edge [pre] node[auto] {} (p4l);
		
		\node [transition] (el) [label = below:$e$, below right = of p4l]{}
		edge [pre] node[auto] {} (p4l);
		
		\node [place] (p5l) [label = below:$p_5$, below right = of dl]{}
		edge [pre] node[auto] {} (dl)
		edge [pre] node[auto] {} (el);
		\end{tikzpicture}
	}
	\centering
	\subfloat{
		\begin{tikzpicture}
		[node distance=0.64cm,
		on grid,>=stealth',
		bend angle=17.5,
		auto,
		every place/.style= {minimum size=2.7mm},
		every transition/.style = {minimum size = 2.9mm}
		]
		\node [place, tokens = 1] (p1) [label = below:$p_1$]{};
		\node[] (N3) [above = of p1]{$N_3$};	
		\node [transition] (a) [label = below:$a$, right = of p1]{}
		edge [pre] node[auto] {} (p1);
		
		\node [place] (p2) [label = below:$p_2$, right = of a]{}
		edge [pre] node[auto] {} (a);
		
		\node [transition] (b1) [label = below:$b_1$, above right = of p2]{}
		edge [pre] node[auto] {} (p2);
		
		\node [transition] (b2) [label = below:$b_2$, below right = of p2]{}
		edge [pre] node[auto] {} (p2);
		
		\node [place] (p3) [label = below:$p_3$, below right = of b1]{}
		edge [pre] node[auto] {} (b1)
		edge [pre] node[auto] {} (b2);
		
		\node [transition] (c) [label = center:$\square$, right = of p3]{}
		edge [pre] node[auto] {} (p3);
		
		\node [place] (p4) [label = below:$p_4$, right = of c]{}
		edge [pre] node[auto] {} (c);
		
		\node [transition] (d) [label = below:$d$, above right = of p4]{}
		edge [pre] node[auto] {} (p4);
		
		\node [transition] (e) [label = below:$e$, below right = of p4]{}
		edge [pre] node[auto] {} (p4);
		
		\node [place] (p5) [label = below:$p_5$, below right = of d]{}
		edge [pre] node[auto] {} (d)
		edge [pre] node[auto] {} (e);
		
		\node [place] (p6) [label = below:$p_6$, above = of c]{}
		edge [pre] node[auto] {} (b1)
		edge [post] node[auto] {} (d);
		
		\node [place] (p7) [label = below:$p_7$, below = of c]{}
		edge [pre] node[auto] {} (b2)
		edge [post] node[auto] {} (e);
		\end{tikzpicture}
	}}
	\caption{$N_2$ is a non-useful refinement and $N_3$ is a useful refinement of $N_1$.}
	\label{fig:split_controlflow}
\end{figure}

Consider event log $L$, relabeling functions $l_1,l_2,l_3$ such that $l_2 \prec l_1 \wedge l_3 \prec l_1$, and event logs $L_1=l_1(L), L_2=l_2(L), L_3=l_3(L)$. Let the $N_1,N_2,N_3$ in Figure~\ref{fig:split_controlflow} be the Petri nets obtained by applying process discovery to $L_1,L_2,L_3$ respectively. The square inside the transition between places $p_3$ and $p_4$ indicates that it is a subprocess.

We can see that refinement $l_2$ does not lead to a meaningful interpretation of $b$ as $b_1$ and $b_2$, since the behavior of the model is not related to the choice between $b_1$ and $b_2$: transitions labeled with $b_1$ and $b_2$ have the same input and output places. Refinement $l_2$ does not provide new insight and unnecessarily harms the understandability of the Petri net by creating more transitions then needed. On the other hand, $l_3$ results in gain of precision, as $\Lan(N_3)$, does not contain $\langle a,b_1,e\rangle$ and $\langle a,b_2,d\rangle$, while $N_1$ does not distinguish between $b_1$ and $b_2$, which suggests that both types of traces are possible.\looseness=-1


\section{Evaluation Method for Label Refinements for Process Models}
\label{sec:calculate_label_split_quality}
In the previous section we showed that we can compare the usefulness of a label refinement by inspecting the Petri net obtained with process discovery. A naive way to evaluate label refinement would be to apply process discovery to all possible label refinements. The number of possible label refinements to consider can however be large and process discovery is a computationally expensive task. Therefore, this naive approach quickly becomes computationally infeasible. We now present a way to estimate the usefulness of a label refinement based on statistics and log relations.

Algorithm \ref{alg:statistical_method} shows the steps of the label refinements evaluation method. The evaluation method consists of an entropy-based component that measures whether a label refinement makes the log statistics more unbalanced, and a statistical test that tests whether there is a label statistic that tests whether the label refinement makes a statistically significant difference to at least one of the log statistics. In the following two sections we described the entropy-based measure and the statistical testing respectively. 

\subsection{Log Statistics}
Event ordering patterns are crucial to most process discovery algorithms. Table \ref{tab:statistics_use} provides an overview of well-known log-based ordering relations described in process discovery literature \cite{Aalst2004,Dongen2004,Wen2006,Weijters2011} and provides examples. Let $L$ be an event log. Let $b,c \in A_L$. Formal definitions of these log-based ordering statistics are as follows:

\begin{itemize}\vspace{-3.04mm}
\item $\#_{L,>}^{+}(b,c)$ is the number of occurrences of $b$ in the traces of $L$ that are directly followed by $c$, i.e. in some $\sigma \in L, i\in\{1,\dots,|\sigma|\}$ we have $[\sigma(i)]_a=b$ and $[\sigma(i+1)]_a=c$ (\emph{direct successor}), $\#_{L,>}^{-}(b,c)$ is the number of occurrences of $b$ which are not directly followed by $c$;
\item $\#_{L,>>}^{+}(b,c)$ and $\#_{L,>>}^{-}(b,c)$ is the number of occurrences of $b$ that are, respectively, are not, followed by $c$: for a trace $\sigma\in L$ and $i\in\{1,\dots,|\sigma|\}$, and $[\sigma(i)]_a=[\sigma(i+2)]_a=b$ and $[\sigma(i+1)]_a=c$ and $b\neq c$ (length-two loops);
\item $\#_{L,>>>}^{+}(b,c)$ and $\#_{L,>>>}^{-}(b,c)$ is the number of occurrences of $b$ that are, respectively are not, eventually followed by $c$: for a trace $\sigma\in L, i,j\in\N$ with $i<j$,  $[\sigma(i)]_a=b$ and $[\sigma(j)]_a=c$ (direct or indirect successor).
\end{itemize}
\vspace{-3.05mm}
In the general sense, let $\#_{L,s}^{+}(b,c)$ and $\#_{L,s}^{-}(b,c)$ be the count of the number of $b$'s that \underline{do}, respectively do not, satisfy relation $s$ in log $L$ with respect to $c$.

\begin{table}
\centering
\caption{Log-based ordering relations and their use by process discovery algorithms}
	\label{tab:statistics_use}
	\scriptsize{
		\begin{tabular}{l p{7cm}}
		\toprule
		Ordering relation & Miners using the relation \\
		\noalign{\smallskip}
		\midrule
		\noalign{\smallskip}
		Direct successor & $a$ miner \cite{Aalst2004}, $a^{++}$ miner \cite{Wen2006}, Multi-phase miner \cite{Dongen2004}, Heuristics miner \cite{Weijters2011} \\
		Length-two loop & $a^{++}$ miner \cite{Wen2006}, Multi-phase miner \cite{Dongen2004}, Heuristics miner \cite{Weijters2011} \\
		Direct/indirect successor & $a^{++}$ miner \cite{Wen2006}, Heuristics miner \cite{Weijters2011} \\
		\bottomrule
		\end{tabular}

	}
\end{table}

\begin{table}
\centering
\caption{A Log statistic in contingency table form}
\label{tab:log_statistics_contingency_table}
	\scalebox{0.85}{
	\begin{tabular}{c|cc|c}
		\toprule
		& $a_1$ & $a_2$ & $a$\\
		\midrule
		$+$& $\#_{L_2,s}^+(a_1,b)$ & $\#_{L_2,s}^+(a_2,b)$ & $\#_{L_1,s}^+(a,b)$ \\
		$-$& $\#_{L_2,s}^-(a_1,b)$ & $\#_{L_2,s}^-(a_2,b)$ & $\#_{L_1,s}^-(a,b)$\\
		\bottomrule
	\end{tabular}}

\end{table} 

\noindent Let $L$ be an event log. Let $l_1$ and $l_2$ be two relabeling functions that are to be compared, such that $l_2 \prec^= l_1$. Let $L_1=l_1(L)$ and $L_2=l_2(L)$. Let $l_1$ and $l_2$ have the property $\{ a_1, a_2 \in A_{L_2}) | \exists_{\sigma_1,\sigma_2 \in L}:l_1(\sigma_1)=\lambda a \wedge l_1(\sigma_2)=\lambda'a \wedge l_2(\sigma_1)=\zeta a_1 \wedge l_2(\sigma_2)=\zeta'a_2 \}\neq\emptyset$, that is, $l_2$ refines activity $a$ into distinct activities $a_1$ and $a_2$. The difference in control flow between $a_1$ and $a_2$ can be expressed as the dissimilarity in log-based ordering statistics between event label $a_1$ and $b \in A_{L_2}\setminus\{a_1,a_2\}$ on the one hand, and $a_2$ and $b$ on the other hand. 
Each log-based ordering statistics of $a_1$ and $a_2$ with regard to any other activity $b$ can be formulated in the form of a contingency table, as shown in Table \ref{tab:log_statistics_contingency_table}.

\subsection{Information Gain}
The binary entropy function, $H_b(p) = -p \log_2 p - (1-p)\log_2(1-p)$, where $0\log_2 0 = 0$, is a measure of uncertainty. Applied on a log statistic, the binary entropy function represents a degree of nondeterminism. Nondeterministic, unbalanced, log statistics are a helpful to process discovery algorithms that operate of log statistics, as it provides low uncertainty to the mining algorithm. Low entropy in the log statistics indicate high predictability of the process, making it easier for process discovery algorithms to return a sensible process model.

Consider the contingency tables in Table \ref{tab:contingency_tables_evaluation}, based on log statistics obtained from Table \ref{tab:evaluation_log} between the events labeled \emph{Tossing \& turning} and \emph{Getting up} and the events labeled \emph{Living room motion}. On the right hand side of the table, separated by the bar, are the log statistics of the before-split label in the before-split log. All five events with label \emph{Getting up} directly precede an event with label \emph{Living room motion}, while all sixteen events with label \emph{Tossing \& turning} are \underline{not} directly preceded by \emph{Living room motion}. Furthermore, all events with refined labels do \underline{not} directly or eventually follow an event with label \emph{Living room motion}, and all events with refined labels do eventually precede an event with label \emph{Living room motion}.

Log statistics with a high degree of non-determinism, like the directly precedes statistic of the bedroom motion events before the split, might confuse a mining algorithm as there is no clear structure here: the \emph{Bedroom motion} event might directly precede \emph{Livingroom motion}, but most of the time it does not. After the split we see a completely deterministic directly precedes statistic, where \emph{Tossing \& turning} never and \emph{Getting up} always directly precedes \emph{Livingroom motion}. This increased determinism is reflected by the entropy of the directly precedes statistic before and after the split. Before the split we have $-\frac{5}{5+16}\log_2\frac{5}{5+16} - \frac{16}{5+16}\log_2\frac{16}{5+16}=0.7919$ bit of entropy in the directly precedes statistic, compared to $-\frac{0}{0+16}\log_2\frac{0}{0+16} - \frac{16}{0+16}\log_2\frac{16}{0+16}=0$ bit of entropy for \emph{Tossing \& turning} and $-\frac{0}{0+5}\log_2\frac{0}{0+5} - \frac{5}{0+5}\log_2\frac{5}{0+5}=0$ bit of entropy for \emph{Getting up}. The conditional entropy of the log statistic after the split is the weighted average of the entropy of the labels created in the split, which is $-\frac{16}{21}0 \times \frac{5}{21}0=0$. 
The information gain of this label split with regard to the directly precedes \emph{Livingroom motion} statistic is equal to the total entropy of the log statistic prior to the split, minus the conditional entropy after the split, this $0.7919-0=0.7919$. Relative information gain \cite{Kullback1951} is a metric that provides insight in the ratio of bits of entropy reduced by a refinement, and can be calculated by dividing the information gain by the before-split entropy. The relative information gain of the directly precedes \emph{Livingroom motion} statistic is $\frac{0.7919}{0.7919}=1$. Figure \ref{fig:precision} shows the effect of this label refinement on the resulting Petri net obtained by process discovery.

So far we have calculated the Relative information gain for a single log statistic. A label refinement however can have impact on multiple log statistics at once. We need a measure that integrates the information gain values of all log statistics to express the quality of a label refinement with respect to the determinism of the log statistics. We therefore sum over the entropy of all log statistics before the label split to obtain the total before-split entropy. We sum over the conditional entropies of all log statistics after the label split to obtain the total after-split entropy. Information Gain and Relative information gain are calculated as before. We let $relative\_information\_gain(L_1,L_2)$ be the function that returns the Relative information gain based on the pre-split log $L_1$ and post-split log $L_2$, where the set of refined label pairs in $L_2$ from which the log statistics are used corresponds to $\{ a_1, a_2 \in A_{L_2} | \exists_{\sigma_1,\sigma_2 \in L}:l_1(\sigma_1)=\lambda a \wedge l_1(\sigma_2)=\lambda^{'} a \wedge$, with the $a$ the corresponding label in $L_1$.

\begin{table}
\scriptsize{
\captionsetup[subfloat]{labelformat=empty}
\caption{Contingency tables for comparing the behavior of the two refined labels}
\label{tab:contingency_tables_evaluation}
\qquad
\centering
\subfloat[][Directly follows]{
\scalebox{0.9}{
\begin{tabular}{p{0.17cm}|p{0.65cm}p{0.64cm}|p{0.6975cm}}
\toprule
& \scriptsize{Tossing \& turning} & \scriptsize{Getting up} & \scriptsize{Bed- room motion} \\
\midrule
$\overset{+}{\rightarrow}$ & 0  & 0 & 0\\
$\overset{-}{\rightarrow}$ & 16 & 5 & 21\\
\bottomrule
\end{tabular}}}
\qquad
\subfloat[][Directly precedes]{
\scalebox{0.9}{
\begin{tabular}{p{0.17cm}|p{0.65cm}p{0.64cm}|p{0.6975cm}}
\toprule
& \scriptsize{Tossing \& turning} & \scriptsize{Getting up} &
\scriptsize{Bed- room motion} \\
\midrule
$\overset{+}{\rightarrow}$ & 0  & 5 & 5\\
$\overset{-}{\rightarrow}$ & 16 & 0 & 16\\
\bottomrule
\end{tabular}}}
\\
\qquad\qquad\qquad
\subfloat[][Eventually follows]{
\scalebox{0.9}{
\begin{tabular}{p{0.17cm}|p{0.65cm}p{0.64cm}|p{0.6975cm}}
\toprule
& \scriptsize{Tossing \& turning} & \scriptsize{Getting up} & \scriptsize{Bed- room motion}\\
\midrule
$\overset{+}{\rightarrow}$ & 0  & 0 & 0\\
$\overset{-}{\rightarrow}$ & 16 & 5 & 21\\
\bottomrule
\end{tabular}}}
\qquad
\subfloat[][Eventually precedes]{
	\scalebox{0.9}{
\begin{tabular}{p{0.17cm}|p{0.65cm}p{0.64cm}|p{0.6975cm}}
\toprule
& \scriptsize{Tossing \& turning} & \scriptsize{Getting up}&
\scriptsize{Bed- room motion} \\
\midrule
$\overset{+}{\rightarrow}$ & 16 & 5 &21\\
$\overset{-}{\rightarrow}$ &  0 & 0 &0\\
\bottomrule
\end{tabular}}}}
\qquad
\vspace{-0.2cm}
\end{table}
\begin{algorithm}
	\scriptsize{
		\begin{tt}
			\noindent Input: Event log $L$, Relabeling functions $l_1$ and $l_2$ such that $l_2 \prec^= l_1$,\\
			Output: the Relative information gain of $l_1$ w.r.t $l_2$,\\
			Parameters: Set of log-based ordering statistics $S$, \\ \indent Significance level $\alpha$.
			\begin{itemize}[topsep=0pt,itemsep=-1ex,partopsep=1ex,parsep=1ex,leftmargin=0cm]
				\item[] all\_significant\_different = true;  $L_1$=$l_1(L)$; $L_2$=$l_2(L)$;
				\item[] split\_set = $\{ a_1, a_2 \in A(L_2) | \exists_{\sigma_1,\sigma_2 \in L}:l_1(\sigma_1)=\lambda a \wedge l_1(\sigma_2)=\lambda^{'} a \wedge$ $l_2(\sigma_1)=\zeta a_1 \wedge l_2(\sigma_2)=\zeta^{'}a_2 \}$; 
				\indentitem \item[] For each $a_1,a_2 \in$ split\_set:
				\indentiitem \item[] $pair\_significant\_different$ = false;
				\indentiitem \item[] For each $\{b \in A(L_2)\setminus\{a_1,a_2\}\}$:
				\indentiiitem \item[] For each $s \in S$:
				\indentiiiitem \item[] p = $fisher\_test( \#_{L_2,s}^{+}(a_1,b), \#_{L_2,s}^{-}(a_1,b),\#_{L_2,s}^{+}(a_2,b), \#_{L_2,s}^{-}(a_2,b))$;
				\indentiiiitem \item[] If($p<\alpha$) pair\_significant\_different = true;
				\indentiitem \item[] If(!$pair\_significant\_different$) 
				\indentiiitem \item[] $all\_significant\_different$ = false;
				\indentitem \item[] If($all\_significant\_different$) \indentiitem \item[] return $relative\_information\_gain(L_1,L_2)$;
				\indentitem \item[] Else return $0.0$;
			\end{itemize}
			\caption{Algorithm of the label refinement statistical evaluation method}
			\label{alg:statistical_method}
		\end{tt}
	}
	\vspace{-0.4cm}
\end{algorithm}
\subsection{Statistical Testing}
Relative information gain can be high by chance for a refinement when the generated refined labels are infrequent. Statistical testing of log statistic differences in addition to calculating relative information gain enables us to distinguish between information gain obtained by chance and actual information gain. Fisher's exact test \cite{Fisher1934} is a statistical significance test for the analysis of contingency tables. When applied to the table above, it calculates a $p$-value for the null hypothesis that $a_1$ and $a_2$ events are equally likely to hold log relation $s$ with regard to label $b$.
Fisher's exact test assumes individual observations to be independent and row and column totals to be fixed. Independence of individual observations might be affected by the grouping of events in traces. In this paper we consider individual observations independence to be working assumption. The test was designed for experiments where both the row and column totals where conditioned. In our setting, the column totals are conditioned by the relabeling function, as the number of events of each label depends on the relabeling. The row totals however, are not conditioned and are an observation. Fisher's exact test is not strictly speaking exact when one or both of the row or column totals are unconditioned, but will instead be slightly conservative \cite{McDonald2009}, meaning that the probability of the p-value being less than or equal to the significance level when the null hypothesis is true is less than the significance level. Fisher's exact test is computationally expensive for large numbers of observations. For large sample sizes, either the $\chi^2$ test of independence or the G-test of independence can be used, which are both found to be inaccurate for small sample sizes. A popular guideline is to not use the $\chi^2$ test of independence or the G-test for samples sizes less than one thousand \cite{McDonald2009}. The computational complexity of the evaluation procedure is $\mathcal{O}(|S|\times|A(L)|\times|split\_set|)$. Many process discovery algorithms are exponential in the number of labels \cite{Aalst2012}. Based on this we can conclude that statistical evaluation of label refinements is computationally less expensive than checking label refinement usefulness through process discovery.

\subsection{Correcting for Multiple Testing}
The computational complexity indicates the number of hypothesis tests performed. When a large set of potential label refinements is evaluated, the evaluation method described is susceptible to the repeated testing problem. The larger the set of hypotheses tested, the higher the probability of incorrectly rejecting the null hypothesis in at least one of the hypothesis tests. Applying a Bonferroni correction \cite{Dunn1959,Dunn1961} to the hypothesis tests performed in the statistical evaluation method of label refinements keeps the familywise error rate constant.
\subsection{Example Case}
\noindent Consider the event log in Table \ref{tab:evaluation_log} and imagine a scenario where a home care worker knows from experience that the elderly always sets his alarm clock at $8$:$30$ AM. Based on such expert knowledge we are able to define a label refinement such that all bedroom movements after $8$:$30$ AM are considered as \emph{Getting up} events, while all other bedroom movements are considered to be \emph{Tossing \& turning} events. The rightmost column shows the refined labels obtained through this expert relabeling function. To evaluate the usefulness of this label refinement from a process model point of view, we apply the statistical evaluation method described in Section \ref{sec:calculate_label_split_quality}. As parameters we set the significance level threshold to the frequently used value of $0.01$.\looseness=-1\\

\begin{wraptable}{r}{3.3cm}
	\hspace{-2cm}
	\vspace{-0.5cm}
	\caption{Results of the statistical tests for the evaluation of label refinement usefulness}
	\scalebox{0.85}{
		\scriptsize{
			\begin{tabular}{*{1}{c}|c}
				\toprule
				Log statistic & P-value \\
				\midrule
				Directly follows & $1$ \\
				Directly precedes & $4.91\times10^{-5}$ \\
				Eventually follows & $1$ \\
				Eventually precedes & $1$ \\
				\bottomrule
			\end{tabular}}}
			\label{tab:statistical_results}
			\vspace{-0.5cm}
		\end{wraptable}
Table \ref{tab:statistical_results} shows the outcome of the statistical tests performed as part of the label refinement usefulness evaluation. Four hypothesis tests have been performed, after Bonferroni correction each hypothesis test is tested at significance level $\frac{0.01}{4}=0.0025$. The direct following statistic of \emph{Tossing \& turning} and \emph{Getting up} with \emph{Living room motion} is statistically significantly given this significance level. The label refinement constructed with expert knowledge is found to be a useful label refinement through statistical evaluation.\looseness-1

\section{Real life evaluation}
\label{sec:evaluation_on_a_case_study}
\begin{figure}[b]
	\vspace{-0.25cm}
	\centering
	\includegraphics[width=\textwidth]{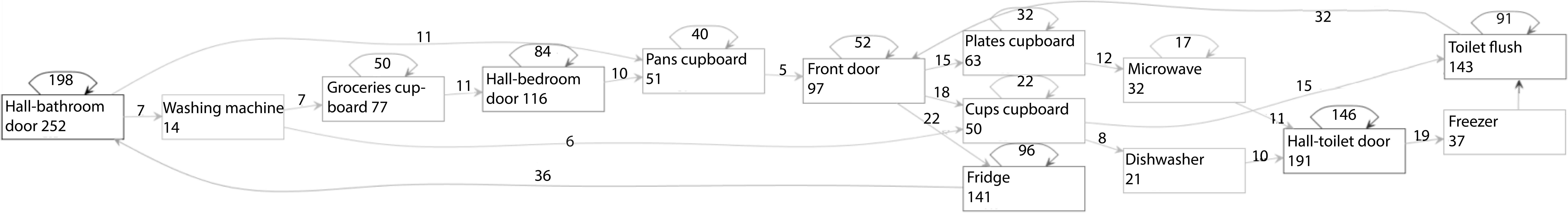}
	\caption{Heuristics net showing original van Kasteren data set}
	\label{fig:kasteren_unrefinement}
\end{figure}

We apply our label refinement evaluation method to a set of candidate label refinements on the Van Kasteren smart home environment data set \cite{Kasteren2008} in order to illustrate the effects of label splits in the context of process mining of real life processes. The van Kasteren data set consists of 1285 events divided over fourteen different sensors. Events are segmented in days from midnight to midnight, to define cases in the event log. The candidate set of label refinements consists of splitting each of the fourteen event types $t$ into two event types based on the their time in the day, such that $t$ events where the time since the start of the day is smaller than the median for $t$ are separated from $t$ events where it is equal to or larger than the median. Figure~\ref{fig:kasteren_unrefinement} shows the \emph{dependency graph} obtained with the Heuristics Miner \cite{Weijters2011}. A \emph{dependency graph} depicts causal relations between activities that meet a certainty threshold. A \emph{dependency graph} can be directly converted into a Petri net \cite{Weijters2011}, however, for the sake of readability we included the \emph{dependency graphs} instead of the Petri nets. The precision \cite{Munoz2010} of the Petri net corresponding to Figure~\ref{fig:kasteren_unrefinement} is 0.56 on a scale from 0 to 1.
\begin{figure}[b]
	\vspace{-0.25cm}
	\centering
	\includegraphics[width=\textwidth]{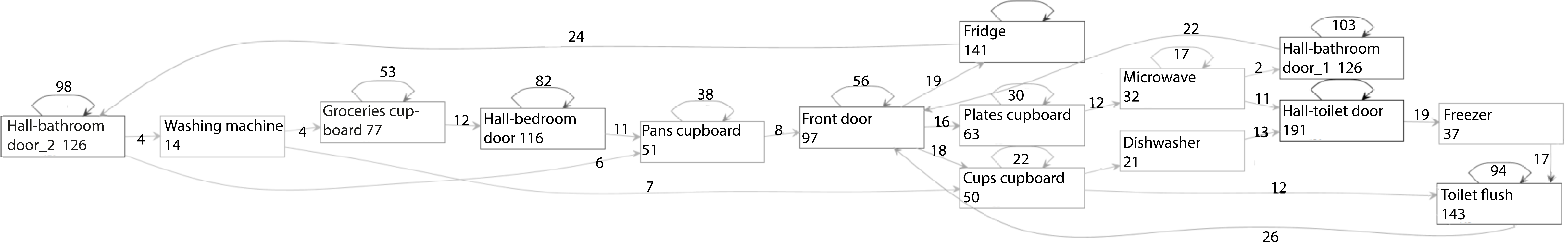}
	\caption{Heuristics net showing the label refinement on \emph{hall-bathroom door} on the van Kasteren data set}
	\label{fig:kasteren_refinement2}
\end{figure} 
\begin{figure}[b]
	\vspace{-0.25cm}
	\centering
	\includegraphics[width=\textwidth]{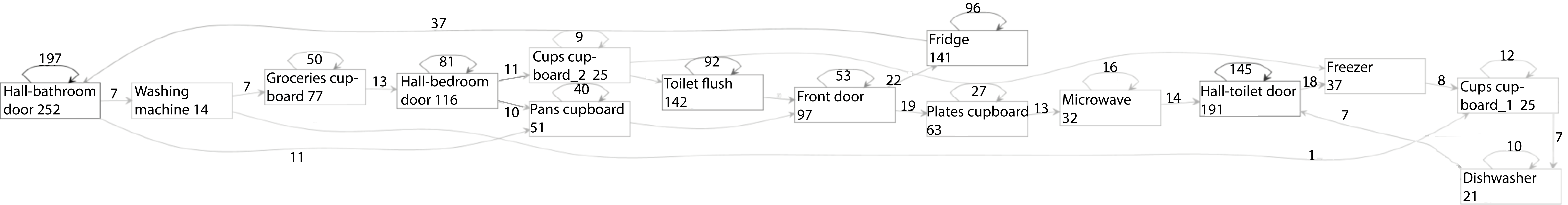}
	\caption{Heuristics net showing the label refinement on \emph{cups cupboard} on the van Kasteren data set}
	\label{fig:kasteren_refinement}
\end{figure}

Out of the fourteen candidate label refinements, two label refinements are selected by our approach. The first label refinement found is the split of \emph{Hall-bathroom door} into \emph{Hall-bathroom door\_1} and \emph{Hall-bathroom door\_2}, with a timestamp below, respectively above or equal to the median time in the day of \emph{Hall-bathroom door} events. The resulting labels of this refinement are statistically significantly different in terms of their eventually follows relation with \emph{Front door} (p-value: $3.06\times10^{-26}$) and their eventually follows relation with \emph{Plates cupboard} (p-value: $3.66\times10^{-23}$) and Microwave \emph{$1.85\times10^{24}$}. The relative information gain on the whole event log caused by this label refinement is 3.47\%. Figure \ref{fig:kasteren_refinement2} shows a Heuristics Net mined with the Heuristics Miner \cite{Weijters2011} on the van Kasteren log with the refined \emph{Hall-bathroom door} label. The model discovered on the log with this label refinement (Figure~\ref{fig:kasteren_refinement2}) has a precision of 0.69, up from 0.53 without the refinement. The increased precision shows that the label refinement helps restricting the share of behavior allowed by the model that is not covered by the event log. The second label refinement found is the split of \emph{Cups cupboard} into \emph{Cups cupboard\_1} and \emph{Cups cupboard\_2}. The resulting labels of this refinement are statistically significantly different in terms of their eventually precedes relation with \emph{Groceries cupboard} (p-value: $2.53\times10^{-34}$) and their eventually follows relation with \emph{Fridge} (p-value: $2.2\times10^{-22}$). The relative information gain on the whole event log caused by this label refinement is 0.53\%. Figure~\ref{fig:kasteren_refinement} shows a Heuristics Net mined with the Heuristics Miner on the van Kasteren log with the refined \emph{Cups cupboard} label, of which the precision is 0.61, up to 0.53 without the refinement. The label refinement with higher information gain also results in a higher improvement in terms of precision, which is in agreement with the intuition that more deterministic log statistics help the miner in mining structured, non-flower-like, models.\looseness=-1

\section{Conclusion \& Future Work}
\label{sec:conclusions}
\vspace{-0.1cm}
We have provided a theoretical and conceptual notion of when label refinements and abstractions are useful from a process discovery point of view. Based on this notion of usefulness, we have shown a framework based on statistics and information theory to evaluate the usefulness of a label refinement or abstraction. In addition, we have shown the applicability of this statistical framework through a real life smart home case, where our method selected two label refinements out of a larger candidate set that increased the precision of the resulting process model. Methods for automatic inference of useful label refinements from event attributes are still to be explored. Such methods may generate a set of candidate label refinements, after which the statistical evaluation method described in this paper can be used to select the most promising label refinement from a set of candidate label refinements.





\bibliographystyle{abbrv}
\bibliography{bibliography}

\begin{thebibliography}{10}

\bibitem{Brank2005}
J.~Brank, M.~Grobelnik, and D.~Mladenic.
\newblock A survey of ontology evaluation techniques.
\newblock In {\em Proceedings of the conference on data mining and data
  warehouses}, pages 166--170, 2005.

\bibitem{Chen2012}
L.~Chen, J.~Hoey, C.~Nugent, D.~Cook, and Z.~Yu.
\newblock Sensor-based activity recognition.
\newblock {\em IEEE Transactions on Systems, Man and Cybernetics Part C:
  Applications and Reviews}, 42(6):790--808, 2012.

\bibitem{Dunn1959}
O.~J. Dunn.
\newblock Estimation of the medians for dependent variables.
\newblock {\em The Annals of Mathematical Statistics}, 30(1):192--197, 1959.

\bibitem{Dunn1961}
O.~J. Dunn.
\newblock Multiple comparisons among means.
\newblock {\em Journal of the American Statistical Association},
  56(293):52--64, 1961.

\bibitem{Fisher1934}
R.~A. Fisher.
\newblock {\em Statistical methods for research workers}.
\newblock Number~5. Genesis Publishing Pvt Ltd, 1934.

\bibitem{Kullback1951}
S.~Kullback and R.~A. Leibler.
\newblock On information and sufficiency.
\newblock {\em The Annals of Mathematical Statistics}, 22(1):79--86, 1951.

\bibitem{Li2007}
J.~Li, D.~Liu, and B.~Yang.
\newblock Process mining: Extending $\alpha$-algorithm to mine duplicate tasks
  in process logs.
\newblock In {\em Advances in Web and Network Technologies, and Information
  Management}, volume 4537 of {\em LNCS}, pages 396--407. Springer Berlin
  Heidelberg, 2007.

\bibitem{Maedche2012}
A.~Maedche.
\newblock {\em Ontology learning for the semantic web}, volume 665.
\newblock Springer Science \& Business Media, 2012.

\bibitem{McDonald2009}
J.~H. McDonald.
\newblock {\em Handbook of biological statistics}, volume~2.
\newblock Sparky House Publishing Baltimore, MD, 2009.

\bibitem{Munoz2010}
J.~Mu\~{n}oz Gama and J.~Carmona.
\newblock A fresh look at precision in process conformance.
\newblock In {\em Business Process Management}, volume 6336 of {\em LNCS},
  pages 211--226. Springer Berlin Heidelberg, 2010.

\bibitem{Quinlan2014}
J.~R. Quinlan.
\newblock {\em C4. 5: programs for machine learning}.
\newblock Elsevier, 2014.

\bibitem{Reisig1998}
W.~Reisig and G.~Rozenberg.
\newblock {\em Lectures on {P}etri nets I: basic models: advances in {P}etri
  nets}, volume 1491.
\newblock Springer Science \& Business Media, 1998.

\bibitem{Rozinat2008}
A.~Rozinat and W.~M.~P. van~der Aalst.
\newblock Conformance checking of processes based on monitoring real behavior.
\newblock {\em Information Systems}, 33(1):64--95, 2008.

\bibitem{Sztyler2015}
T.~Sztyler, J.~V{\"o}lker, J.~Carmona, O.~Meier, and H.~Stuckenschmidt.
\newblock Discovery of personal processes from labeled sensor data--an
  application of process mining to personalized health care.
\newblock In {\em Proceedings of the International Workshop on Algorithms \&
  Theories for the Analysis of Event Data, ATAED}, pages 22--23, 2015.

\bibitem{Aalst2011}
W.~M.~P. van~der Aalst.
\newblock {\em Process mining: discovery, conformance and enhancement of
  business processes}.
\newblock Springer Science \& Business Media, 2011.

\bibitem{Aalst2012}
W.~M.~P. van~der Aalst.
\newblock Distributed process discovery and conformance checking.
\newblock In {\em Fundamental Approaches to Software Engineering}, LNCS, pages
  1--25. Springer, 2012.

\bibitem{Aalst2004}
W.~M.~P. van~der Aalst, A.~J. M.~M. Weijters, and L.~Maruster.
\newblock Workflow mining: Discovering process models from event logs.
\newblock {\em IEEE Transactions on Knowledge and Data Engineering},
  16(9):1128--1142, 2004.

\bibitem{Dongen2004}
B.~F. van Dongen and W.~M.~P. van~der Aalst.
\newblock Multi-phase process mining: Building instance graphs.
\newblock In {\em Conceptual Modeling--ER 2004}, volume 3288 of {\em LNCS},
  pages 362--376. Springer, 2004.

\bibitem{Kasteren2008}
T.~van Kasteren, A.~Noulas, G.~Englebienne, and B.~Kr{\"o}se.
\newblock Accurate activity recognition in a home setting.
\newblock In {\em Proceedings of the 10th International Conference on
  Ubiquitous Computing}, pages 1--9. ACM, 2008.

\bibitem{Broucke2014}
S.~K. Vanden~Broucke.
\newblock {\em Advanced in Process Mining: Artificial Negative Events and Other
  Techniques}.
\newblock PhD thesis, KU Leuven, 2014.

\bibitem{Weijters2011}
A.~J. M.~M. Weijters and J.~T.~S. Ribeiro.
\newblock Flexible heuristics miner (fhm).
\newblock In {\em Proceedings of the 2011 IEEE Symposium on Computational
  Intelligence and Data Mining}, pages 310--317. IEEE, 2011.

\bibitem{Wen2006}
L.~Wen, J.~Wang, and J.~Sun.
\newblock Detecting implicit dependencies between tasks from event logs.
\newblock {\em Frontiers of WWW Research and Development-APWeb 2006}, pages
  591--603, 2006.

\end{thebibliography}
\end{document}